# Exciton-Exciton Interaction and Cascade Relaxation of Excitons in Colloidal CdSe Nanoplatelets

Aleksandr M. Smirnov,[†,¶] Anastasiya D. Golinskaya,[†,§] Bedil M. Saidzhonov,[‡] Vladimir N. Mantsevich,[\*,†,§] Vladimir S. Dneprovskii,[†] and Roman B. Vasiliev[‡]

[†]*Chair of Semiconductors and Cryoelectronics, Physics Department, Lomonosov Moscow State University, 119991 Moscow, Russia*

[‡]*Materials Science Department, Lomonosov Moscow State University, 119991 Moscow, Russia*

[¶]*Kotelnikov Institute of Radioengineering and electronics of RAS, 125009 Moscow, Russia*

[§]*Quantum Technology Center, Physics Department, Lomonosov Moscow State University, 119991 Moscow, Russia*

E-mail: vmantsev@gmail.com

## Abstract

We experimentally investigated for the first time the lowest four band structure of CdSe nanoplatelets at the $\Gamma$ point of Brillouin zone [heavy-hole (hh), light-hole (lh) and two spin-orbital bands] and its modification due to the changing of the CdS shell thickness. The results of an experimental and theoretical study of exciton dynamics in colloidal CdSe/CdS nanoplateles are presented in the case of stationary laser excitation. The excitation was performed by means of the powerful nanosecond laser and the pump and probe technique was utilized. The differential transmission spectra peculiarities were experimentally revealed and allowed to demonstrate and to understand the role of exciton-exciton interaction and phonon-induced cascade relaxation of free electrons and holes in the lower exciton state considering subsequent radiative recombination. The theoretical explanation of exciton-exciton interaction and exciton cascade relaxation was performed.

# Introduction

Semiconductor nanocrystals attract a great deal of attention due to their unique physical and chemical properties, which are very promising for opto-electronical applications.[1] Nanocrystals can be prepared by colloidal synthesis, which enables the creation of nanocrystals with a high degree of control under their shape,[2] size[3] and crystal structure.[4] Recently, colloidal semiconductor quantum wells (nanoplatelets) have been synthesized. Colloidal semiconductor nanoplatelets (NPLs) are atomically flat systems, which demonstrate a zinc-blend crystal structure with a [001] axis. They reveal strong quantum confinement as NPLs are strongly anisotropic systems with several nanometers thick and tens of nanometers in lateral dimensions,[5] which can be used to tune the optical absorption and photoluminescence spectra.[6-9] Moreover, NPLs exhibit great exciton binding energy about several hundreds of meV,[10,11] which makes excitonic resonances in these systems visible at room temperature.[12-14] They also demonstrate intriguing optical properties such as narrow emission lines at both low and room temperatures, tunable emission wavelength, short radiative lifetimes, giant oscillatory strength, high quantum yield and the absence of inhomogeneous broadening.[15-18] All these properties have made NPLs very attractive for application in optoelectronic devices as bright and flexible light emitters,[19-21] lasers,[10,22] biomedical labeling[23] and polarized emitters.[24] Despite a set of potential applications of colloidal NPLs, experimental analysis of the fundamental properties of these systems was mostly concentrated on the different time ranges of the excited states dynamics, such as decay pathways of the single-exciton state,[11,25] recombination dynamics of band edge excitons,[18] non-radiative Auger recombination[10,26] or photoluminescence decay dynamics.[27,28]

It was found that exciton-exciton interaction plays an important role,[10,26] which suggests pronounced non-linear optical effects.[29–32] Although various NPLs heterostructures have been investigated, non-linear optical properties of these structures have not been studied yet in details to the best of our knowledge.

Here, we report results of the experimental observation and theoretical analysis of excitons dynamics in colloidal CdSe/CdS NPLs with different shell thickness under stationary excitation. Using the two color pump probe technique we observed peculiarities in the differential transmission spectra which allowed us to reveal the properties of exciton-exciton interaction and to investigate the cascade relaxation of excitons. We discussed the physics for these effects.

The paper is organized as follows: In Sec. II we describe the procedure of colloidal synthesis of NPLs and their structural characterization. Then in Sec. III we discuss experimental procedure and experimental results, which are theoretically analyzed in Sec. IV. Comparison of theoretical model and experimental results allowed us to estimate parameters of exciton dynamics. The results are summarized in Sec. V.

## II. Synthesis and characterization of nanopleletes

We have studied core-shell NPLs based on thin population of CdSe NPLs with 463 nm lowest energy absorption band having 3.5 ML thickness according to Ref.[33,34]

Synthesis of 3.5 ML CdSe NPLs: 0.5 mmol of cadmium acetate, 0.2 mmol of oleic acid (OA) and 10 mL of 1-octadecene (ODE) were loaded into a reaction flask. The mixture was degassed under stirring and argon flow at $1870^oC$ for 30 min. Then the temperature was increased to $210^oC$ and 150 $\mu L$ of 1 M trioctilphosphine selenide solution diluted in 350 $\mu L$ of ODE was injected into the flask. After 40 min of the growth at $240^oC$, the reaction was quenched by adding 1mL of OA and the mixture was rapidly cooled to room temperature.

As-grown NPLs were washed twice with acetone. Finally, the NPLs were dispersed in 6 ml of hexane.

Preparation of CdSe/NCdS NPLs:[35] 1 mL of hexane dispersion of CdSe NPLs was mixed with 1 mL of 0.1 M $N_2S$ solution in N-methylformamide (NMF) and stirred for 10 min. Then, the solution was left to react for 30 min. After that, the solution was washed twice with acetonitrile/toluene mixture. Afterwards, 1 mL of 0.3 M cadmium acetate solutuion in NMF was added to the precipitate, and the resulting mixture was stirred for 10 min. After 50 min, the mixture was washed twice using acetonitrile and toluene. Finally, the core-shell NPLs were dispersed in 1 mL of NMF. The procedure described above corresponds to the growth of a single CdS ML. To obtain CdSe/NCdS heterostructures, this process was repeated N times [2 ML of CdS shell on each basal plane to get 582 nm absorption core/shell nanoparticles (CdSe/CdS582), 3 ML of CdS shell on each basal plane to get 603 nm absorption core/shell nanoparticles (CdSe/CdS603) and 4 ML of CdS shell on each basal plane to get 615 nm absorption core/shell nanoparticles (CdSe/CdS615)].

A LEO912 AB OMEGA system with an accelerating voltage of 100 kV was used to obtain transmission electron microscope (TEM) images of the samples (see Fig. 1). Samples for TEM measurements were prepared by evaporating one drop of NPL dispersions onto carbon-coated copper grids. Low-resolution transmission electron microscopy (TEM) image of as-grown CdSe/NCdS NPLs demonstrate two-dimensional morphology. Rectangle platelets are flat with lateral sizes about 30 × 100 nm. Well-defined rectangle CdSe/NCdS platelets with same lateral sizes as CdSe NPLs are clearly observed.

# III. Measurements procedure and experimental results

The nonlinear and differential transmission spectra were measured at room temperature in the colloidal solution of CdSe/CdS NPLs in the case of stationary excitation. Scheme of the experimental setup is shown in Fig.2a. Concentration of NPLs in the liquid solution of NMF was about $10^{15}$ $cm^{-3}$. Pumping was realized by the Q-switched $Nd^{3+}$:$YAlO_3$ laser second harmonic ($\lambda = 540$ nm, pulse duration about 10 ns). As a probe we used a broadband photoluminescence radiation of the Coumarin-120, Coumarin-7 and Kiton Red dyes excited by the third harmonic ($\lambda = 360$ nm) of the laser.[36] The probe pulse duration was 11 ns, and we adjusted it to overlap with the pump pulse. The broad spectrum of the probe light is shown in Fig. 2b, it covers wavelengths from 410 to 510 nm by Coumarin-120 photoluminescence, from 480 to 600 nm by Coumarin-7 photoluminescence and from 590 to 660 nm by Kiton Red photoluminescence. It allowed to analyze nonlinear absorption changing for $1hh - 1e$, $1lh - 1e$ and two spin-orbital exciton transitions. Pump intensities vary from 0.2 $MW/cm^2$ to 9 $MW/cm^2$ during the experimental measurements. Pump intensity tuning was realized by means of the neutral optical filters. The transmission and absorption of the probe light was measured with spectral resolution using SpectraPro 2300$i$ spectrometer with PIXIS 256 CCD-camera.

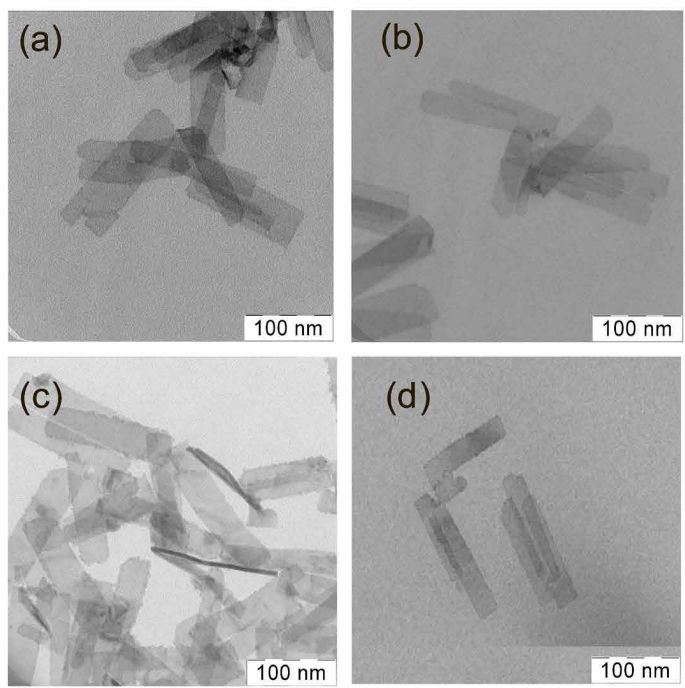

Figure 1: Low-resolution transmission electron microscopy (TEM) overview images of as-synthesized CdSe/Cds NPLs. (a) CdSe/CdS540 NPLs with 1 ML of CdS on each basal planes; (b) CdSe/CdS582 NPLs with 2 ML of CdS on each basal planes; (c) and CdSe/CdS603 NPLs with 3 ML of CdS on each basal planes; (d) and CdSe/CdS615 NPLs with 4 ML of CdS on each basal planes.

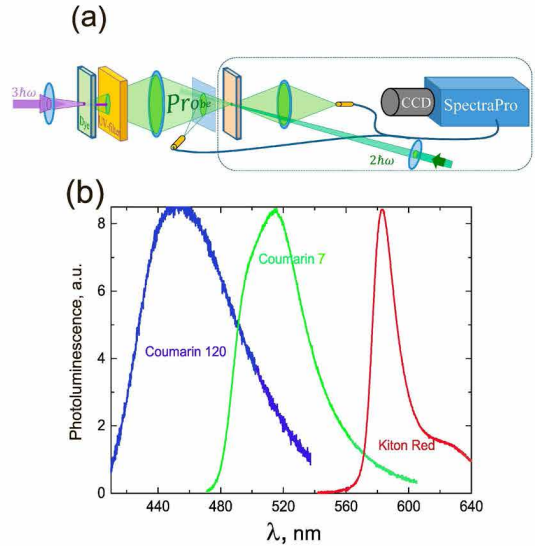

Figure 2: (a) Scheme of the experimental setup used for pump and probe measurements. Dashed contoured part of the experimental setup was used to measure the PL spectra; (b) Spectrum of the probe light.

Linear (black) and nonlinear (red) transmission spectra of colloidal NPLs samples in 1 mm cell with NMF are shown in Fig.3. We revealed an increase in transmission for all NPLs colloidal solutions. Analogous to bulk semiconductor CdSe, NPLs have the conduction band, the heavy hole band and the light hole band.[25] It leads to the presence of two well-resolved distinct dips in the transmission spectra (see Fig. 3). The high (low) energy dip is due to electronic excitation from the light (heavy) hole valence band state to the conduction band state. One could also resolve smooth high energy dips which could be associated with two $1so - 1e$ exciton transitions. The energies of these transitions are governed by the quantum confinement influence, as the thickness of the CdSe layer is smaller than the exciton Bohr radius in the bulk CdSe.

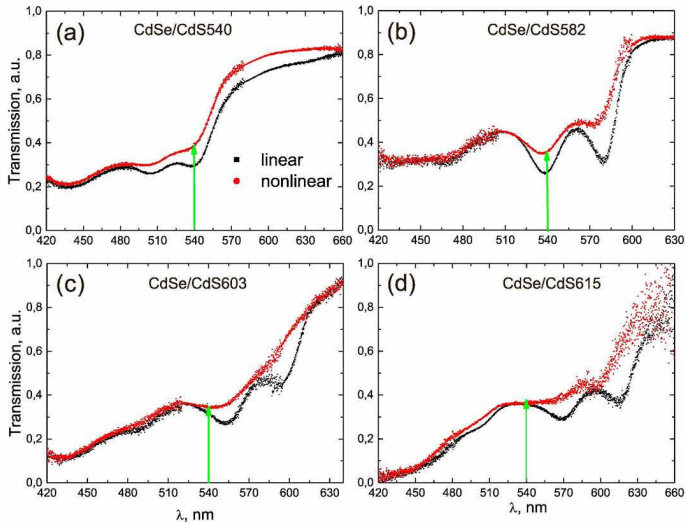

Figure 3: (Color online) Absorption spectra of four different colloidal NPLs samples. Excitation was performed at 8.6 $MW/cm^2$. Green arrow shows the pump laser wavelength.

The differential transmittance spectra for the colloidal solutions of NPLs are presented in Fig.4. The differential transmission at a given wavelength $\lambda$ is

$$DT(\lambda) = \frac{T_I(\lambda) - T_0(\lambda)}{T_0(\lambda)}, \qquad (1)$$

where $T_I(\lambda)$ is the transmission of the solution of colloidal NPLs excited by the laser pulses and $T_0(\lambda)$ is the linear transmission.

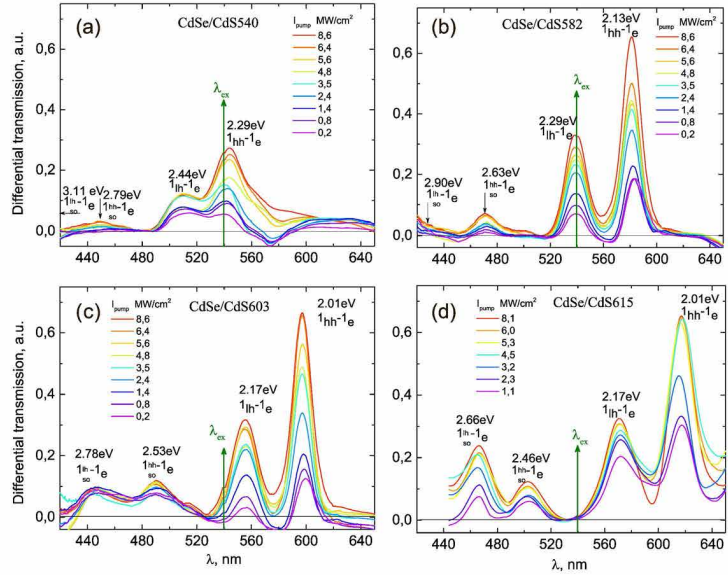

Figure 4: (Color online) The measured differential transmittance spectra for the different pump laser pulses intensity.

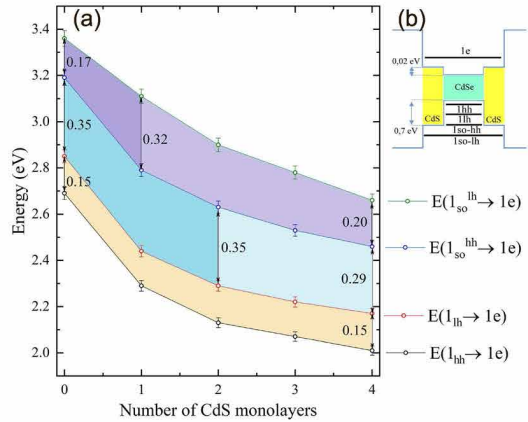

Figure 5: (Color online) (a) Experimentally measured exciton transition energies of colloidal CdSe/CdS NPLs with different shell thickness. (b) The scheme of the CdSe/CdS NPLs energy levels.

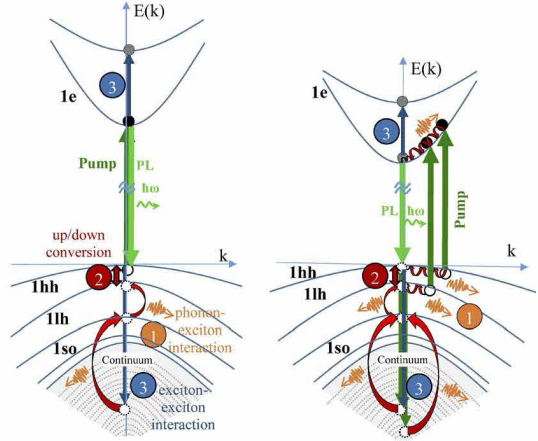

Figure 6: (Color online) Schemes of the excitons excitation and relaxation in CdSe/CdS NPLs. (1) demonstrates the processes of exciton dephasing caused by interaction with phonons, (2) demonstrates energy exchange between heavy-hole and light-hole excitons and (3) outlines the process of exciton-exciton interaction with the formation of free electrons and holes; red arrows indicate relaxation of carriers through the quasi-continuous hole states and the thermalization with the participation of phonons.

## IV. Discussion

Changing the thickness of the NPLs shell one can resonantly or non-resonantly excite light hole or heavy hole excitons. Differential transmission spectra of all studied samples demonstrate a significant growth of transmission of exciton transitions from the heavy-hole and light-hole sub-bands to the conduction band ($1hh - 1e$ and $1lh - 1e$) with pump intensity increasing. The measured splitting between heavy holes and light holes sub-bands in the colloidal CdSe/CdS NPLs $\Delta E(1lh - 1hh) = 150 \pm 5$ meV was found to be independent on the CdS shell thickness (see Fig.5a). This result can be explained by the holes bands localization in the NPLs core associated with the deep quantum well - the difference between the tops of the valence bands in CdS and CdSe is about 0.7 eV (see Fig.5b).[37] Electrons are delocalized in the core and shell for CdSe/CdS nanocrystals[38] - the difference between the bottoms of the conduction bands is around 20 meV (see Fig.5b). The width of the quantum well for the electrons increases with the growth of the shell thickness, which causes the decrease of the $1hh - 1e$ and $1lh - 1e$ transition energies (see Fig.5a).

One can also resolve two additional transitions in room temperature differential trans-mission spectra at the energies higher than the heavy-hole and light-hole exciton transitions occur (see Fig.4). These two transitions do not depend on the excitation regime and are clearly seen for both resonant and non-resonant excitation of heavy-hole and light-hole excitons. We associate these two high energy peaks with the absorption changing caused by the electron-hole transitions corresponding to the spin-orbital sub-band $1so^{hh}-1e$ and $1so^{lh}-1e$, which was theoretically predicted in.[39] They correspond to the E symmetry of wave functions at the $\Gamma$ point. A large spin-orbit splitting for E bands originates from the states at the upper part of the valence band. The energy splitting between these two sub-bands decreases with increasing of the NPLs shell thickness (number of shell monolayers). Such behavior indicates that these sub-bands are delocalized in the core and in the shell of the CdSe/CdS NPLs (see Fig.5b).

The observed bleaching of the heavy- and light-hole excitons was explained by the phase space filling effect in the case of single-photon excitation. The number of excitons in each NPL increases with the growth of the excitation intensity. It results in the exciton phase space filling effect and, consequently, causes the reduction of the NPLs colloidal solution absorption. To reveal the contribution of the phase space filling effect in the colloidal solution nonlinear absorption reduction with the growth of excitation intensity, the number of excitons in each NPL corresponding to the saturation concentration was estimated. Considering the area of NPL $S_{NPL}$ and the saturation parameter $N_S = \frac{7}{32}\frac{1}{\pi a_{2D}^2}$[43] one can estimate the number of excitons to be about $\eta = S_{NPL}N_S \simeq 20$. Two-dimensional exciton radius is about 30.[44]

To understand the saturation processes one should consider the relaxation mechanism of excited carriers. The relaxation of both resonantly and non-resonantly excited excitons in the NPLs (see Fig.6) includes the following processes: radiative recombination involving heavy-hole exciton (light green arrow in Fig.6); the excitons dephasing processes, which occur through the excitons interaction with acoustic and optical phonons (circle 1 in Fig.6); the energy exchange between light-hole and heavy-hole excitons (circle 2 in Fig.6).[34] The excitons relaxation to the state with minimal energy occurs at the sub-picosecond times,[40,41] which are much smaller than the duration of excitation laser pulses.

The revealed absorption decreasing of the excitons transitions corresponding to the spin-orbital sub-bands can be explained by the processes of exciton-exciton interaction (circle 3 in Fig.6) and subsequent rapid picosecond relaxation through deep quasi-continuous hole bands.[45] The excitons interaction leads to the acceleration of relaxation and to the appearance of the electron-hole plasma in NPLs. The characteristic relaxation times for exciton-exciton interaction are about 10 ps,[26] which is significantly smaller than the duration of the exciting pulses. The absorption dependence on the excitation intensity changes significantly when a number of excited excitons exceeds the saturation density $N_S$. The exciton-exciton interaction starts to play a significant role.[42] Otherwise this process is also called two-exciton Auger recombination.[26,44] For the analysis of the recombination and interaction of excitons in the colloidal nanoplatelets under high excitation densities, one should consider the rate equations describing the exciton-exciton and exciton-electron interactions. At room temperatures the rate of excitons generation differs from the rate of electron-hole pairs generation $G(t)$. Simple rate equations for exciton ($N_{exc}$) and electron ($N_{el}$) concentrations read:

$$\begin{aligned}
\frac{dN_{exc}}{dt} &= \gamma_{exc}N^2 - \frac{N_{exc}}{\tau_{exc}} - \alpha N_{exc}^2, \\
\frac{dN_{el}}{dt} &= G(t) - (\gamma_{el} + \gamma_{exc})N_{el}^2 - \frac{N_{el}}{\tau_{el}} + \frac{1}{2}\alpha N_{exc}^2.
\end{aligned} \quad (2)$$

Parameters $\tau_{exc}$ and $\tau_{el}$ characterize the lifetime of the excitons due to their decay and the electron relaxation time, respectively. $\gamma_{el}$ describes direct recombination of electrons and $\gamma_{exc}$ corresponds to the recombination via excitons and $\alpha$ is the constant, which characterises exciton-exciton interaction processes. The function $G(t)$ is taken in the form of the rectangular pulse of duration $\tau = 10^{-8}$s and the amplitude $10^{26}$cm$^{-3}$s$^{-1}$ to $10^{28}$ cm$^{-3}$s$^{-1}$, which corresponds to our experimental conditions. To estimate constant $\alpha$ one can use the value $10^{-8}$cm$^{-3}$s$^{-1}$ to $10^{-6}$cm$^{-3}$s$^{-1}$.[42] The electron relaxation time $\tau_{el} \simeq 10^{-10} \div 10^{-11}$ s[46] and the radiative exciton lifetime $\tau_{exc} \simeq 10^{-10} \div 10^{-9}$s. We estimate $\gamma_{exc}$ to vary from $\gamma_{exc} = 0.1(\gamma_{exc} + \gamma_{el})$ to $\gamma_{exc} = (\gamma_{exc} + \gamma_{el})$, the latter corresponds to the case of recombination only through the exciton states. The coefficient $\gamma_{exc}$ was taken to be $10^{-7}$ sm$^{-3}$s$^{-1}$ to $10^{-6}$ sm$^{-3}$s$^{-1}$. Modeling results and calculation parmeters are presented in Fig.7. The break separates the stage of initial growth of electrons and excitons concentration directly after the laser pulse and the relaxation stage after switching off the laser pulse. Changing of electrons and excitons concentration demonstrates that the maximum of excitons concentration is shifted relative to the concentration of electrons. Electrons and holes bounds to the excitons directly after the laser pulse reaches the colloidal solution. The decreasing of excitons concentration after switching off the laser pyulse occurs due to the cascade relaxation processes.

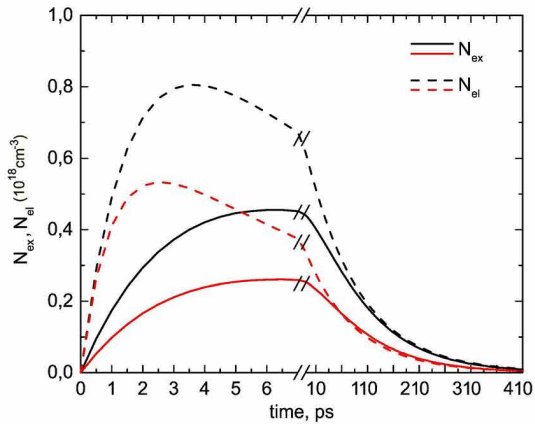

Figure 7: Time evolution of the electron (dashed curves) and exciton (solid curves) concentration in the presence of exciton-exciton interaction. The break separates the stage of the initial growth of electrons and excitons concentration directly after the laser pulse and the relaxation stage after switching off the laser pulse. The parameters used for the calculation are: $G(t) = 10^{28}$ cm$^{-3}$s$^{-1}$, $\tau_{el} = 3 \times 10^{-11}$s, $\tau_{exc} = 5 \times 10^{-10}$s, $\gamma_{exc} + \gamma_{el} = 10^{-7}$ cm$^{-3}$s$^{-1}$. For black curves $\alpha = 10^{-7}$ cm$^{-3}$s$^{-1}$ and for red curves $\alpha = 3 \times 10^{-6}$ cm$^{-3}$s$^{-1}$.

# V. Conclusion

We for the first time experimentally analyzed the lowest four band structure of colloidal core/shell CdSe/CdS NPLs at the $\Gamma$ point of Brillouin zone. Analyzing differential transmis-sion spectra we revealed the presence of heavy-hole, light-hole and two spin-orbital sub-bands and investigated their modification due to the changing of the CdS shell thickness. The results of an experimental and theoretical study of exciton dynamics in colloidal CdSe/CdS NPLs are obtained in the regime of the stationary laser excitation. The revealed differential transmission spectra peculiarities allowed to demonstrate and to understand the role of exciton-exciton interaction and phonon-induced cascade relaxation of free electrons and holes in the lower exciton state considering subsequent radiative recombination.

# Acknowledgement

We acknowledge the support by the Russian Science Foundation (Project $18-72-10002$ all experimental measurements and theoretical calculations) and the RFBR grant $19-03-00481$ (samples synthesis).

# Graphical TOC entry

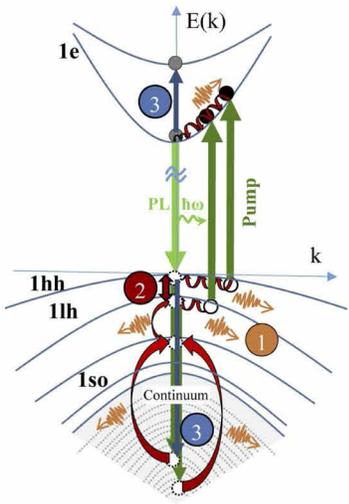